# Distance measurement in air without the precise knowledge of refractive index fluctuation

**Morris Cui[1], Steven A. van den Berg[2], Nandini Bhattacharya[1]**

[1]Optics Research Group, Delft University of Technology, the Netherland
[2]VSL, Thijsseweg 11, 2629 JA Delft, The Netherlands
*Corresponding author: n.bhattacharya@tudelft.nl

## Abstract

The accuracy of long distance measurement in air is limited by the fluctuation of refractive index. In this paper, we propose a technique which allows us to measure an absolute distance in air without the knowledge of air turbulence. The technique is based on a femtosecond frequency comb. The fluctuation of the environmental conditions is monitored by two independently operating reference interferometers. The deviations of optical path lengths, caused by the fluctuation of air refractive index, is compensated by feedbacks from the reference interferometers. The measured optical path length is then locked to certain environmental conditions, determined at an optimized moment before the measurement process.

Laser beam propagation in air is subject to refractive index of air. Applications where this is especially relevant are laser based distance measurements. For most laboratory measurements under controlled environments this can be managed, though the accuracy of the measurement depends upon the determination of refractive index. The air refractive index, for most laser based distance measurements is calculated from empirical equations, which have an uncertainty of $3\times10^{-8}$ [1, 2]. These empirical equations themselves need data of the environmental parameters which then determine the instantaneous refractive index. In general an uncertainty of 0.1 K in air temperature or 0.4 hPa in pressure cause an uncertainty of $10^{-7}$ in the calculated value of the refractive index of air at the moment of measurement. Monitoring the refractive index of air becomes even more difficult when either the measurement distance is long or the measurement duration is long. On the other hand field measurements have to work around the additional burden of turbulence where the spatial and temporal fluctuation of the air refractive index becomes a serious contributor to the uncertainty of the measurement.

Even with these constraints the techniques for laser based distance measurements have gone through a quiet revolution in the last decade. Technology advances in optical sources, electronics and improved computational facilities have led to higher accuracy and repeatability of measurements [3]. Many applications ranging from projection based manufacturing, geodetic monitoring, interferometry in space to fundamental physics experiments based on interferometry are driving the progress in these field by increasing the demands on dimensional metrology. In case the application demands that the measurements are performed in vacuum for instance in the case of space based interferometers or the vacuum tubes of particle physics experiments, the main factors influencing performance are the stability of the optical source and measurement noise. Other applications can have challenging environments like indoor or outdoor measurements or automated manufacturing but still require low measurement uncertainties.

Laser based ranging for long distances in air are now reaching relative uncertainties of $10^{-6}$ for distances which are hundreds of meters in length [4, 5]. This is made possible mainly by stabilized pulsed laser sources which have a comb like frequency spectrum. The stabilization techniques used by this laser source transfers the accuracy of time standards, with relative uncertainties that can be as low as $10^{-14}$ for a cesium atomic clock to the frequency domain by connecting the microwave frequencies to optical frequencies. Even with these developments the fundamental problem of measurement in air remains the inaccuracy arising from the determination of refractive index.

Distance measurement in air using optical frequency combs has been demonstrated using many techniques. There is comb referenced interferometry [6, 7], direct pulse interferometry [8, 9, 10], spectral interferometry [11, 12, 13], time of flight [14], dual comb [5, 15] just to name a few. Most of these measurements report accuracies of the order of $10^{-8}$ and are limited by the determination of refractive index. Using multi-color length measurements [16, 17] which compensate for refractive index of air the accuracies are similar since the exact refractive index determination is the predominant bottleneck.

Refractive index change during a measurement due to turbulence or change of environmental conditions may also lead to additional uncertainty on the distance to be measured. In this work, we propose a technique which allows us to measure an absolute distance in air beyond the knowledge of air fluctuation. The measurement is based on using a femtosecond frequency comb as the illumination source. With this technique the optical path length is always locked to a specific environmental condition even though the environmental condition is fluctuating. Two independent reference interferometers continuously monitor the fluctuation of the refractive index of air. The optical path length deviation caused by this fluctuation is compensated by a piezo element in the measurement interferometer.

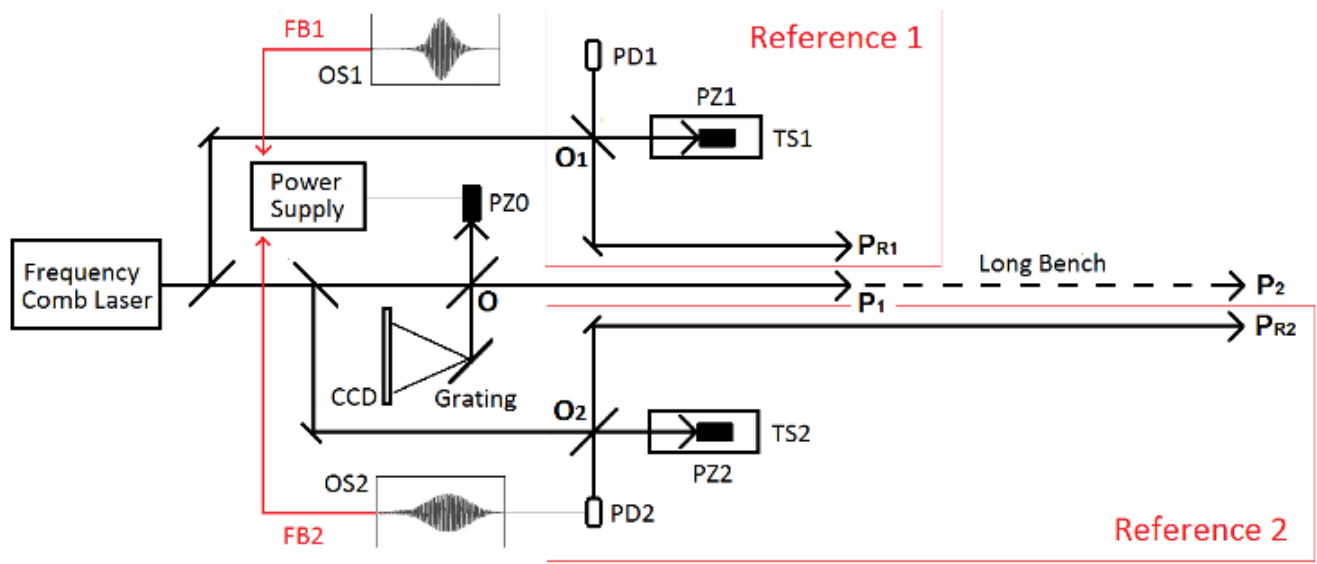

Fig. 1. Schematic of the experimental setup. The light beam from the frequency comb is sent into three interferometers. The measurement interferometer is used to measure an optical path length between $P_1$ and $P_2$. The two reference interferometers are used to monitor the fluctuations of environmental conditions and provide instantaneous feedback to the measurement interferometer.

The frequency comb source is a mode-locked pulsed laser, with both the repetition frequency $f_r$ and the carrier-envelop frequency $f_0$ locked to a cesium atomic clock. The light beam from the frequency comb is separated into three interferometers: a measurement interferometer and two reference interferometers.

Fig. 1 shows the schematic of the setup. The measurement interferometer is used to measure an optical path length at a certain locked environmental condition at a reference moment $t_0$. It is a dispersive interferometer, with the short arm consisting of a retro-reflector, mounted on a piezo electric actuator (PZ0). The position of the piezo is controlled by the feedback, provided by the two reference interferometers. The retro-reflector of the long arm is mounted on a mechanical car which can move along the long bench. $P_1$ and $P_2$ and indicate the positions of the initial and final position of the measurement for a specific environmental condition. The interference signal of the measurement interferometer is dispersed by a grating spectrometer. The spectral interferogram is detected and recorded by a CCD line camera.

The first reference interferometer (Reference 1) is used to monitor the air fluctuations between $O$ and $P_1$, and provides instantaneous feedback to PZ0 during the measurement at $P_1$. It is a Michelson interferometer, with the short arm periodically scanned by a piezo element (PZ1). The length of the short arm can be changed by a translation stage (TS1). The long arm of Reference 1 is co-collimated with $OP_1$. The end of the long arm, $P_{R1}$, does not necessarily overlap, but must be close enough to $P_1$. Hence, $O_1P_{R1}$ shares the same air refractive index with $OP_1$. The interference of the two arms is detected by a photodiode (PD1). The cross-correlation is observed by an oscilloscope (OS1).

The second reference interferometer (Reference 2) is used to monitor the air fluctuations between $O$ and $P_2$, and provides instantaneous feedback to PZ0 during the measurement at $P_2$. It has a similar configuration as Reference 1, with the long arm now collimated with $OP_2$. The interference is detected by PD2, and the cross-correlation is observed by OS2.

Before the distance measurement, the two reference interferometers should be aligned. Move the two translational stages, TS1 and TS2, until two cross-correlations are observed in OS1 and OS2. The positions of the cross-correlations are floating, due to the fluctuations of air refractive index along $O_1P_{R1}$ and $O_2P_{R2}$. Hence the instant positions of the cross-correlations are representatives of certain environmental conditions of the paths $O_1P_{R1}$ and $O_2P_{R2}$. Subsequently the two cross-correlations in OS1 and OS2 are recorded. The time of recording the two cross-correlations is labelled as $t_0$. The two recorded cross-correlations carry the signature of the environmental parameters *(T, p, h)* at $t_0$ and thus the optical path length. Both $P_{R1}$ and $P_{R2}$ are fixed until all measurements are done.

In an absolute distance measurement, the mechanical car is first positioned at $P_1$. The spectral interferogram at $P_1$ is observed by the CCD line camera of the measurement interferometer. The spectral interferogram is floating due to the fluctuation of the refractive index due to the constantly varying environmental parameters. The cross-correlations at OS1 are continuously monitored and compared to the cross-correlation recorded at $t_0$. This comparison is used to generate the feedback, FB1, which can be then fed to the piezo electric actuator PZ0. With the feedback on, the spectral interferogram at $P_1$ is stabilized because the optical path length deviation, caused by the fluctuation in $OP_1$, due to the environmental parameters, is compensated by the displacement of PZ0. The compensated spectral interferogram is then recorded and $t_1$ is the time stamp given to the recording moment.

In the next step FB1 is switched off and the mechanical car is displaced along the bench to $P_2$, the new position. The floating spectral interferogram $P_2$ can be observed. Using the same protocol as used for position $P_1$, and using OS2 to generate the instantaneous feedback, FB2, for PZ0 the second compensated spectral interferogram is recorded and the recording time is labelled as $t_2$.

The distance $|P_1P_2|$ can be calculated using the relation [11],

$$|P_1P_2| = m \cdot L_{pp}/2 + L_1/2 - L_2/2 \qquad (1)$$

Here, $m$ is an integer number and $L_{pp}$ is the inter-pulse distance, calculated from the repetition frequency $f_r$ by,

$$L_{pp} = c/(n_g f_r) \qquad (2)$$

where $n_g$ is the group refractive index of air at $t_0$. $L_1$ and $L_2$ are distances between the pulses from the two arms of the measurement interferometer at $P_1$ and $P_2$ respectively. $L_1$ and $L_2$ can be obtained from the spectral interferograms at $P_1$ and $P_2$ by,

$$L_{1,2} = c/(n_{air} \Delta f_{1,2}) \qquad (3)$$

where $\Delta f$ is the distance between adjacent peaks on the interferogram, an example can be seen in Fig.2 (c) and $n_{air}$ is the refractive index of air at $t_0$.

In the proposed technique, the refractive index at $t_0$ is the only parameter we need during the whole measurement. We can choose $t_0$ at an optimized time when temperature, pressure and humidity can be carefully controlled, or under quiet weather conditions. The air condition before and after $t_0$ can all be ignored. In principle, the time interval between $t_0$ and $t_1$ can be arbitrarily long, although mechanical shifts of the retroreflectors at $P_{R1}$ and $P_{R2}$ should be taken into consideration. The correlations recorded at $t_0$ can be multiple used, and the measured distances using the same locking conditions are comparable.

In the case when $P_1$ is close to $O$ and the measurement duration at $P_1$ is short, the air fluctuation at $P_1$ during the measurement is ignorable, hence Reference 1 is not necessary. In such a case, $t_0$ can be chosen at the same time of the measurement at $P_1$, and only the air fluctuation along $OP_2$ needs to be compensated.

A numerical simulation was performed to explain the technique. The simulation uses a Gaussian shaped pulse, centered at 800 nm ($3.747 \times 10^{14}$ Hz), with a FWHM of $1 \times 10^{13}$ Hz (~ 20 nm). The repetition rate of the laser was taken to be 1 GHz. For the reference time, $t_0$, the environmental parameters were taken to be temperature, $T = 20.0$ °C, pressure, $p = 1013.25$ hPa and relative humidity $RH = 40\%$ in the simulation. . Fig. 2 shows the result of the simulation for the proposed measurement at $P_{R2}$ and $P_2$. The cross-correlation observed at $t_0$ in OS2 is shown in Fig. 2 (a). In this case, the path-length difference of the reference interferometer (Reference 2) is 66 cavity lengths (approximately 20 m). The cross-correlation at $t_2$ is shown in Fig. 2(b). In the simulation, we change the temperature at $t_2$ to 20.5 °C, and keep $p$ and $RH$ the same as at $t_0$. The 0.5 °C in temperature change brings 9.4 µm shift of the cross-correlation. This shift is used as compensation signal to PZ0. (c) is the detected spectral interferogram without the correction from Reference 2. (d) is the spectral interferogram with the correction loop from RF2. With the feedback loop, the spectral interferogram at $P_2$ is locked to the environmental condition of $t_0$. The fringe density on (d) is smaller than on (c), manifesting a smaller optical path length difference ($L_2$) after the self-correction of air refractive index.

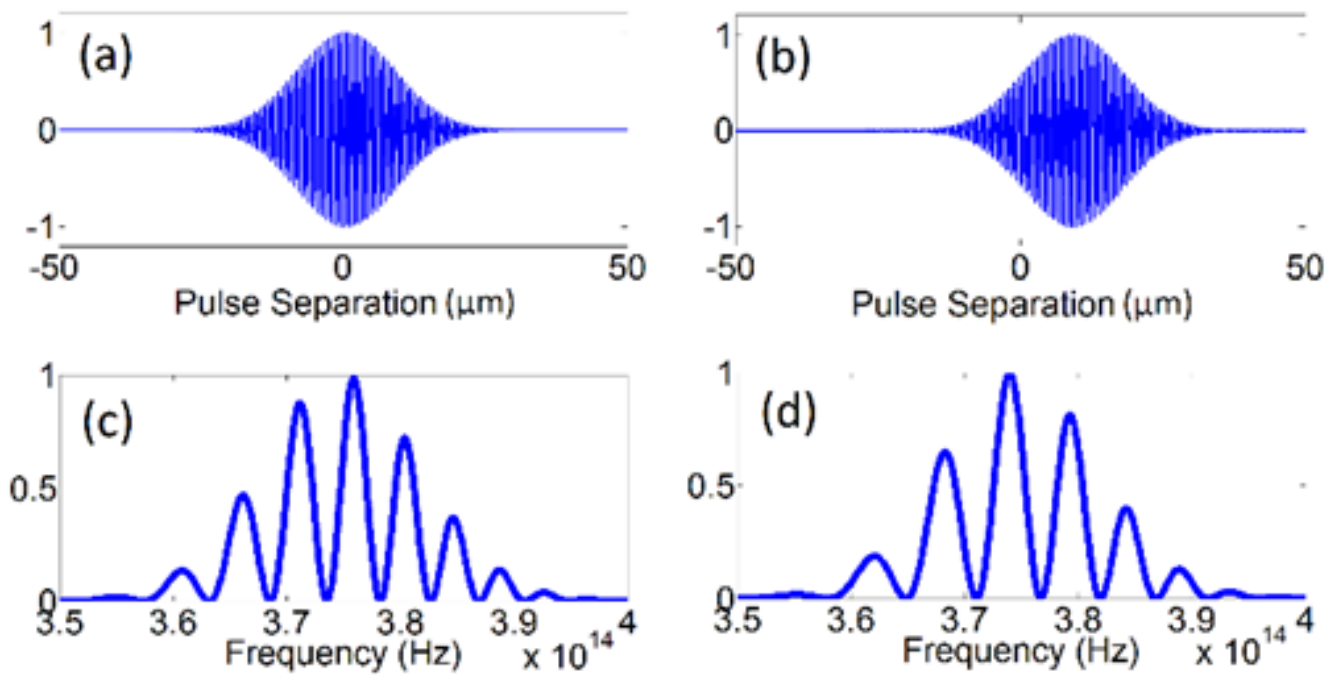

Fig. 2. Simulated results of the measurement at the points $P_{R2}$ and $P_2$, at the long arm of the measurement interferometer (a) cross-correlation measured by PD2 at $t_0$. (b) cross-correlation measured by PD2 at $t_2$. (c) spectral interferogram at $P_2$, without feedback from Reference 2. (d) spectral interferogram at $P_2$, with feedback from Reference 2.

Our design is especially useful for the cases when temporal or spatial fluctuation of air is strong, such as in outdoor environment [18]. Since the reference interferometers provide instant feedbacks to the measurement interferometer, temporal fluctuations of the air refractive index are instantly corrected. Because $O_1P_{R1}$ and $O_2P_{R2}$ are collimated with $OP_1$ and $OP_2$, the spatial change of the refractive index on the measurement interferometer is also observed by the reference interferometers.

Our proposed experimental setup can also be used in industrial systems where the precise optical distance has to be maintained between components or projection systems. In such systems, without the self-correction of $n_{air}$, the small mechanical shift of one component will be immersed into the uncertainty of the refractive index of air, hence cannot be detected. With the proposed system, the optical path length change due to refractive index of air is cancelled, hence the mechanical change between the two points stands out. When we only want to monitor the mechanical displacement at $P_2$, the knowledge of absolute refractive index of air at $t_0$ is not needed.

At last, our setup has potential usage to create an "artificial" table environment. Some experiments need an especially stable environment which is too difficult or too expensive to be achieved in laboratory condition. For example, some space research need environment of long distance in vacuum. Building a long distant vacuum tunnel is expensive and difficult. In such cases, our setup can be used to mimic stable environment, which provide conditions to make experiments be done.